\documentclass{osa-article}

\journal{oe}

\articletype{Research Article}

\usepackage{braket}
\usepackage{amsmath}

\usepackage{siunitx}
\usepackage{ulem}
\usepackage{bbm}
\usepackage{color}
\usepackage{soul}
\definecolor{mygreen}{rgb}{0,0.5,0}
\definecolor{mygrey}{rgb}{0.5,0.5,0.5}
\definecolor{myred}{rgb}{0.75,0,0}
\definecolor{myblue}{rgb}{0,0,0.75}
\definecolor{mymagenta}{cmyk}{0,1,0,0.12}
\definecolor{mycyan}{cmyk}{1,0,0,0.12}
\definecolor{myorange}{rgb}{1,0.5,0}
\definecolor{myviolet}{rgb}{0.5,0.0,0.75}

\definecolor{mybrown}{rgb}{0.542969,0.269531, 0.0742188} 

\newcommand{\commentout}[1]{}

\newcommand{\var}{\mathrm{var}}

\newcommand{\supplus}{^{(\rm L)}}

\newcommand{\suppm}{^{(\rm L,R)}}

\newcommand{\subin}{_{\rm in}}
\newcommand{\supint}{^{\rm int}}
\newcommand{\subT}{_{\cal T}}
\newcommand{\subR}{_{\cal R}}
\newcommand{\phiat}{\phi_\mathrm{at}}
\newcommand{\hphiat}{\hat{\phi}_\mathrm{at}}

\newcommand{\subcavone}{_{\rm C1}}
\newcommand{\subcavtwo}{_{\rm C2}}

\newcommand{\subwindow}{_{\rm W}}
\newcommand{\subatoms}{_{\rm A}}
\newcommand{\subprop}{_{\rm L}}
\newcommand{\subsys}{_{\rm sys}}
\newcommand{\supSP}{^\mathrm{SP}}
\newcommand{\supin}{^\mathrm{in}}
\newcommand{\supout}{^\mathrm{out}}
\newcommand{\supCE}{^\mathrm{CE}}

\address{}

\begin{document}

\newcommand{\mytitle}{Cavity-enhanced atomic polarization rotation measurements }
\renewcommand{\mytitle}{Cavity-enhanced polarization rotation measurements for low-disturbance probing of atoms  }

\title{\mytitle}

\newcommand{\ICFO}{ICFO - Institut de Ciencies Fotoniques, The Barcelona Institute of Science and Technology, Castelldefels (Barcelona) 08860, Spain}
\newcommand{\ICREA}{ICREA - Instituci\'{o} Catalana de Recerca i Estudis Avan{\c{c}}ats, 08010 Barcelona, Spain}

\author{Chiara Mazzinghi\authormark{1*},
Daniel Benedicto Orenes\authormark{1}, Pau Gomez\authormark{1}, Vito G. Lucivero\authormark{1}, Enes Aybar\authormark{1}, Stuti Gugnani\authormark{1} 
and Morgan W. Mitchell\authormark{1,2}}

\address{\authormark{1}\ICFO}
\address{\authormark{2}\ICREA}

\email{\authormark{*}chiara.mazzinghi@icfo.eu} 


\begin{abstract}
We propose and demonstrate  cavity-enhanced polarization-rotation measurement as a means to detect magnetic effects in transparent media with greater sensitivity at equal optical disturbance to the medium.  Using the Jones calculus, we compute the effective polarization rotation effect in a Fabry-Perot cavity containing a magnetic medium, including losses due to enclosure windows or other sources. The results show that when measuring polarization rotation, collecting the transmitted light has advantages in simplicity and linearity relative to collecting the reflected light. We demonstrate the technique by measuring Faraday rotation in a $^{87}$Rb atomic ensemble in the single-pass and cavity-enhanced geometries, and observe enhancement in good agreement with the theoretical predictions. We also demonstrate shot-noise-limited operation of the enhanced rotation scheme in the small-angle regime.
\end{abstract}

\section{Introduction}
Dispersive measurement of atomic variables using near-resonant light is both a practical method of non-destructive probing \cite{KominisN2003s,GriffithOE2010s} and a versatile technique for preparing non-classical states of matter \cite{JulsgaardN2001, TakanoPRL2009, SewellPRL2012}.  When the atomic system is contained within an optical resonator, both the measurement uncertainty and the disturbance to the measured variable can be far below the intrinsic quantum noise, leading to strong squeezing and entanglement \cite{Schleier-SmithPRL2010, BohnetNPhot2014, HostenN2016}.  To date, most cavity-enhanced probing techniques have been applied to non-magnetic transitions in atomic systems, of interest for atomic clocks \cite{LerouxPRL2010b, ValletNJP2017}. There has been less work with cavity-enhanced measurement of atomic magnetization \cite{CrepazSR2015,chang2015cavity,lucivero2021femtotesla}, of interest to magnetometers \cite{Budker2007}, gyroscopes \cite{KornackPRL2005}, and instruments to search for physics beyond the standard model \cite{LeePRL2018, gomez2020,safronova2018search}.  

Polarization rotation (PR) measurements, in which a probe beam's linear polarization rotates upon propagating through a medium, are widely applied to detect magnetic effects in atomic media.  These include Faraday rotation, in which the magnetic field produces circular birefringence via the Zeeman effect \cite{BudkerRMP2002}, and the so-called paramagnetic Faraday rotation, in which the spin polarization or magnetization of the medium is responsible for the rotation \cite{TakahashiPRA1999}. While it is to be expected that PR measurements can be enhanced by the Purcell effect created by cavity resonance, the methods used to date, including transmission \cite{Schleier-SmithPRL2010, BohnetNPhot2014} and Pound-Drever-Hall (PDH) \cite{HostenN2016, ValletNJP2017}
measurements of the line shift of a single cavity mode, are not directly applicable to PR, which involves two modes of different polarization. This intrinsically two-mode character of PR probing presents novel challenges in the design of cavity-enhancement (CE) methods.

In this work we propose and demonstrate a novel cavity-enhancement scheme to boost the sensitivity of PR measurements. We first describe the atom-light system using the Jones matrix formalism \cite{jones1941, hechtoptics} to obtain analytic expressions for the cavity's output field in transmission and in reflection. We note various inconveniences of a PDH-like PR measurement in reflection, and focus on the transmission geometry.  We then report an experimental implementation, in which a resonant optical cavity is built around a  \textsuperscript{87}Rb vapor cell. We compare the polarization rotation angle with and without the cavity, and observe an enhancement factor in agreement with the theoretical model. We measure the noise arising in this implementation, and conclude that the probing system is shot-noise limited in a relevant regime for pulsed probing of cold atomic ensembles \cite{Sewell2012, Colangelo2017}.
The new probing scheme will enable cavity-enhanced quantum-non-demolition  measurement \cite{SewellNP2013} of magnetic degrees of freedom in hot and cold atomic systems, creation of non-classical states of matter \cite{Colangelo2017, KongNC2020}, and magnetic sensitivity beyond the standard quantum limit \cite{MartinPRL2017}.

\begin{figure}[t]
	\centering
	\includegraphics[scale=0.7]{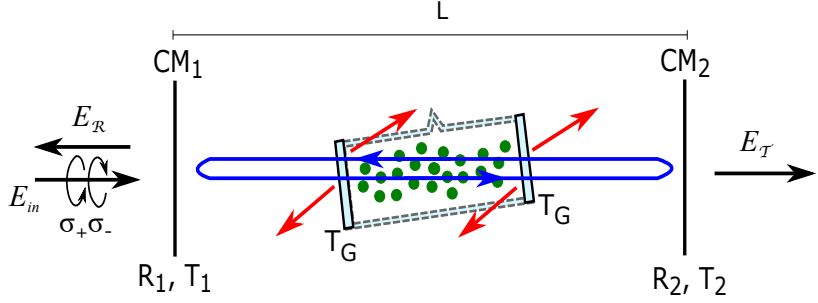}
	\caption{\textbf{Model for cavity-enhanced polarization-rotation measurements.} A quasi-monochromatic field $E\subin$ impinges upon an optical cavity formed by mirrors CM$_1$ and CM$_2$, with intensity reflection (transmission) coefficients $R_1 (T_1)$ and $R_2 (T_2)$, respectively. An atomic medium in the cavity introduces a phase shift $\mp \phiat$ for the $\sigma_\pm$ polarizations on each pass (blue oval) between cavity mirrors. Optical windows with intensity transmission coefficients $T_G$ scatter light out of the cavity mode (red arrows), but do not produce back-reflection.  Reflected or transmitted output field $E_{\cal R}$ or $E_{\cal T}$, respectively, is analyzed by measurement of the Stokes parameter $S_2$. } 
	\label{fig:cavity_diagram}
\end{figure}

\section{Model of cavity-enhanced polarization rotation}
The components of the model are illustrated in \autoref{fig:cavity_diagram}: an atomic medium is placed within a Fabry-Perot optical resonator that, apart from possible birefringent effects arising from the atomic medium itself, is polarization-independent. Horizontally polarized input light, with electric field $E\subin$, impinges upon the cavity and output fields $E\subR$ (reflected) and $E\subT$ (transmitted), which are in general elliptically polarized, containing information about the magnetic conditions within the atomic medium. A balanced polarimeter (not shown in \autoref{fig:cavity_diagram}) detects a Stokes component of $E\subR$ or $E\subT$.

\subsection{Jones Matrix description}

We treat the probe light as quasi-monochromatic, and assume it is spatially matched to the cavity's fundamental TEM$_{00}$ mode. At any given point in space and in a given direction (forward or backward propagating), the field is completely specified by the amplitudes of circular left (L) and circular right (R) polarizations. Because the cavity \textit{per se} is polarization-insensitive and the atoms are not, it is more convenient to describe the field in terms of its amplitudes for $\sigma_+$ and $\sigma_-$ polarization, i.e. the circular polarizations with respect to the atomic quantization axis, chosen to be the forward propagation direction.

\newcommand{\subloc}{_\mathrm{\alpha}}
\renewcommand{\supplus}{^{(+)}}
\newcommand{\supminus}{^{(-)}}
\renewcommand{\suppm}{^{(\pm)}}
\newcommand{\identitymatrix}{\left(\begin{array}{cc} 1 & 0 \\ 0 & 1 \end{array} \right) }
\renewcommand{\identitymatrix}{\mathbbm{1}}
\newcommand{\zeromatrix}{\left(\begin{array}{cc} 0 & 0 \\ 0 & 0 \end{array}\right)}
\renewcommand{\zeromatrix}{0 \cdot\mathbbm{1}}

We thus define Jones vectors $E\subloc \equiv (E\subloc\supplus,E\subloc\supminus)^T$, where $\alpha$ indicates a location and direction, and $E\subloc\suppm$ indicates the electric field amplitude for the $\sigma_\pm$ polarization at $\alpha$.  For example, the input field, immediately before reaching the input mirror and propagating in the forward direction, is $E_\mathrm{in}=E_0 \ (1/\sqrt{2},1/\sqrt{2})^T$, where $E_0$ is the amplitude of the input field. Each optical transformation, e.g. transmission through the medium, propagation by a distance, or reflection from a mirror, is represented by a matrix that acts on the Jones vector. These matrices are given in \autoref{tab:JonesMatrices}.

\begin{table}[t]
    \centering
\begin{tabular}{c|c|c}
\hline \hline
 optical transformation & symbol &  Jones matrix  \\
 \hline 
free space propagation (round trip)     &  $P_\mathrm{L}$ & $e^{i\phi_0} \identitymatrix$\\
transmission through mirror $m$    &  $T_{\mathrm{C}_m}$ & $\sqrt{T_m} \identitymatrix$\\
reflection from mirror $m$    &  $R_{\mathrm{C}_m}$ & $\sqrt{R_m} \identitymatrix$\\
transmission through window    &  $T_{\mathrm{W}}$ & $\sqrt{T_G} \identitymatrix$\\
reflection from window  &  N/A & $\zeromatrix$ \\
transmission through medium    &  $F_A$ & $\left(\begin{array}{cc} e^{-i \phiat/2}  & 0 \\ 0 & e^{i \phiat/2}  \end{array}\right)$\\
\hline \hline
 polarization component & Stokes component &  detection matrix  \\
 \hline 
total power      &  $S_0$ & $\frac{1}{2}\identitymatrix$\\
linear      &  $S_1$ & $\frac{1}{2}\left(\begin{array}{cc} 0  & 1 \\ 1 & 0 \\
\end{array}\right)$\\
diagonal      &  $S_2$ & $\frac{1}{2}\left(\begin{array}{cc} 0  & -i \\ i & 0 \\
\end{array}\right)$\\
\end{tabular}  
\caption{Jones matrices used to describe various steps of field propagation in the model. $\mathbbm{1}$ indicates the $2\times 2$ identity matrix.  $\phi_0$ is the phase shift acquired on a round trip of the cavity, including polarization-independent phases acquired in propagation through the atomic medium, cell windows, and upon reflection from mirrors. $R_m$ and $T_m$ are the intensity reflection and transmission coefficients of the $m$th mirror.  $T_G$ is the intensity transmission coefficient of the windows, assumed equal. Reflection from a window is assumed to fall outside the TEM$_{00}$ mode, and thus does not propagate further within the model. $\phiat$ is the relative phase between $\sigma_+$ and $\sigma_-$ polarizations upon passing through the atomic medium, equal to half the geometrical polarization rotation angle. Because the PR is of magnetic origin, the same matrix $F_A$ applies for either forward or backward propagation through the medium.   }
\label{tab:JonesMatrices}
\end{table}

\newcommand{\subeff}{_{\rm eff}}

It is convenient to use Stokes components \cite{stokes,hechtoptics} to describe the polarization at the detection stage. Considering a pulse of duration $\tau$ and effective beam cross-section $A\subeff$, the Stokes parameter $S_\beta$, in photon-number units, is 
\begin{equation}
\label{eq:StokesFromE}
{S}_\beta = \frac{c\tau A\subeff \epsilon_0}{4} E_{\rm out}^{*} \hat{S}_\beta E_{\rm out},
\end{equation} where $\hat{S}_\beta$ is the corresponding matrix, given in \autoref{tab:JonesMatrices}. Considering that the input light is fully polarized in the horizontal direction, i.e.,  $S_1\supin = S_0\supin$, the angle of polarization rotation is $\psi = \arcsin({S}_2\supout/S_0\supout)$. The superscript ``in'' refers to the forward-propagating field that enters the system, and ``out'' refers to the output field, which could be in the forward or backward direction, depending on where the detection is placed. 

\subsection{Single pass and cavity schemes}
\label{Sec:theory}
We consider three cases: single pass (i.e. without cavity), cavity in reflection and in transmission. We will be interested in two figures of merit, the rotation gain $\kappa \equiv \psi/\phiat$ and the system efficiency $\eta\subsys \equiv S_0\supout / S_0\supin$. 

\subsection{Single pass PR measurement}
\label{sec:SPth}
The output field is
\begin{equation}
E_\mathrm{SP}
=T\subwindow F\subatoms  T\subwindow E\subin = \frac{E_0 T_G}{\sqrt{2}} 
\begin{pmatrix}
e^{-i \phiat/2} \\
e^{i \phiat/2}
\end{pmatrix}
\end{equation}
from which $ \psi\supSP =\phiat$ and thus $\kappa\supSP = 1$. The system efficiency is $\eta\subsys\supSP =T_G^2$.

\subsection{Cavity enhanced PR measurement}
\label{sec:cavityTh}

We repeat the calculation of the output field as in previous section, but include now the cavity mirrors.  The input field is considered as in the previous section, and the output field  $E\subT$ ($E\subR$) from cavity in transmission (in reflection) is:
\begin{eqnarray}
E\subT&=&
T\subcavone T\subcavtwo T\subwindow^2 P\subprop F\subatoms \sum_{p=0}^{\infty}\Bigl(R\subcavone R\subcavtwo T\subwindow^4 P\subprop^2 F\subatoms^2 \Bigr)^p E\subin \label{eq:Etransm} \\
E\subR&=&
\biggl[-R\subcavone + T\subcavone^2 R\subcavtwo T\subwindow^4 P\subprop^2 F\subatoms^2 \sum_{p=0}^{\infty}\Bigl(R\subcavone R\subcavtwo T\subwindow^4 P\subprop^2 F\subatoms^2 \Bigr)^p\biggr]  E\subin.
\label{eq:Ereflection}
\end{eqnarray}
The geometric series can be evaluated analytically, to find
\begin{eqnarray}
\label{eq:ETransPM}
E_{\mathcal{T}}^{(\pm)}&=&\frac{E_0}{\sqrt{2}} \biggl(\frac{\sqrt{T_1 T_2} T_G \exp{[i \phi_0 \pm i\phiat/2]}}{1- T_G^2 \sqrt{R_2 R_1} \exp{[i \phi_0 \pm i\phiat]}} \biggr) \\
\label{eq:EReflPM}
E_{\mathcal{R}}^{(\pm)}&=&\frac{E_0}{\sqrt{2}} \biggl(-\sqrt{R_1} +\frac{\sqrt{R_2} T_1 T_G^2 \exp{[i \phi_0 \pm i\phiat]}}{1- T_G^2 \sqrt{R_2 R_1} \exp{[i \phi_0 \pm i\phiat]}} \biggr)
\end{eqnarray}
For simplicity and since it is the highest-sensitivity scenario, in what follows we consider only the on-resonance case, that is $\phi_0 = 0$.   
To quantify the performance of the cavity in boosting the rotation signal, we define
 \begin{equation}
 \label{eq:psicavat}
 \psi\supCE \equiv \arcsin({S_2^{\mathcal{T},\mathcal{R}}}/ {S_0^{\mathcal{T},\mathcal{R}}} )
 = 
 \kappa_{\mathcal{R,T}} \psi\supSP
 \end{equation}
where $\kappa_{\mathcal{R,T}}$ is the cavity enhancement factor, i.e. rotation gain factor. 
The system efficiency is $\eta\subsys^{{\cal R},{\cal T}} \equiv S_0^{\mathcal{T},\mathcal{R}} / S_0\supin$, in analogy with the single-pass case. 

\newcommand{\subphot}{_{\rm phot}}

\subsection{Sensitivity and sensitivity enhancement at equal disturbance}
The shot-noise limited sensitivity to rotation angle can be quantified using the propagation of error formula. Writing $\hphiat$ for the estimate of $\phiat$, the mean squared error (MSE) of $\hphiat$ is 
\begin{eqnarray}
\mathrm{MSE}(\hphiat) &=& \left(\frac{\partial S_2\supout}{\partial \phiat}\right)^{-2}\var(S_2\supout) \nonumber \\
\label{eq:MSE}
& \approx & \left(\frac{S_0\supout \partial \psi}{\partial \phiat}\right)^{-2} \frac{S_0\supout}{2},
\end{eqnarray}
where the approximation assumes $\psi\ll 1$ and that shot noise dominates over electronic and technical noise. 

\newcommand{\subint}{_{\rm int}}

Meanwhile, for linear-optical disturbance effects such as optical pumping,  the disturbance to the atoms is proportional to the spatially-averaged probe intensity. 
Defining as $2 S_0\supint$ the photon number seen by the atoms, $S_0\supint$ can be expressed in terms of $S_0\supin$, which through $\eta\subsys$ is related also to $S_0\supout$. In the SP case $(S_0\supint)_{\rm SP}=T_G (S_0\supin)_{\rm SP}$. In the CE case, the forward-propagating intracavity field $E\subint$ is 
\begin{align}
    E\subint&=
T\subcavone T\subwindow \sum_{p=0}^{\infty}\Bigl(R\subcavone R\subcavtwo T\subwindow^4 P\subprop^2 F\subatoms^2 \Bigr)^p E\subin,
\end{align}
Assuming $\phiat\ll 1$ and $\phi_0=0$, i.e., cavity resonance,  evaluating again analytically the geometric series, using \autoref{eq:StokesFromE} and multiplying by two to account for the fact that the intra-cavity light passes the atoms both in the forward and backward directions, we obtain 
\begin{align}
(S_0\supint)_{\rm CE}&=\frac{2 T_1 T_G}{(1-\sqrt{R_1 R_2}T_G^2)^2} (S_0\supin)_{\rm CE}
\end{align}

For a comparison at equal disturbance, we take $(S_0\supint)_{\rm CE} = (S_0\supint)_{\rm SP}$, which implies the relation  $(S_0\supin)_{\rm CE} = (1-\sqrt{R_1 R_2}T_G^2)^2 \ (S_0\supin)_{\rm SP}/(2 T_1)$ between the input powers.
Using \autoref{eq:MSE}, the CE versus SP MSE ratio at equal disturbance is then

\begin{equation}
    \frac{\mathrm{MSE}(\hphiat)_{\rm CE}}{\mathrm{MSE}(\hphiat)_{\rm SP}}=\frac{2 \eta\subsys^{\rm SP} T_1}{\eta\subsys^{{\cal R},{\cal T}} (1-\sqrt{R_1 R_2}T_G^2)^2 \kappa_{\mathcal{R,T}}^2} \equiv \zeta_{\cal R,T}.
    \label{Eq:zeta}
\end{equation}

\begin{figure}[t]
    \centering
    \includegraphics[scale=0.5]{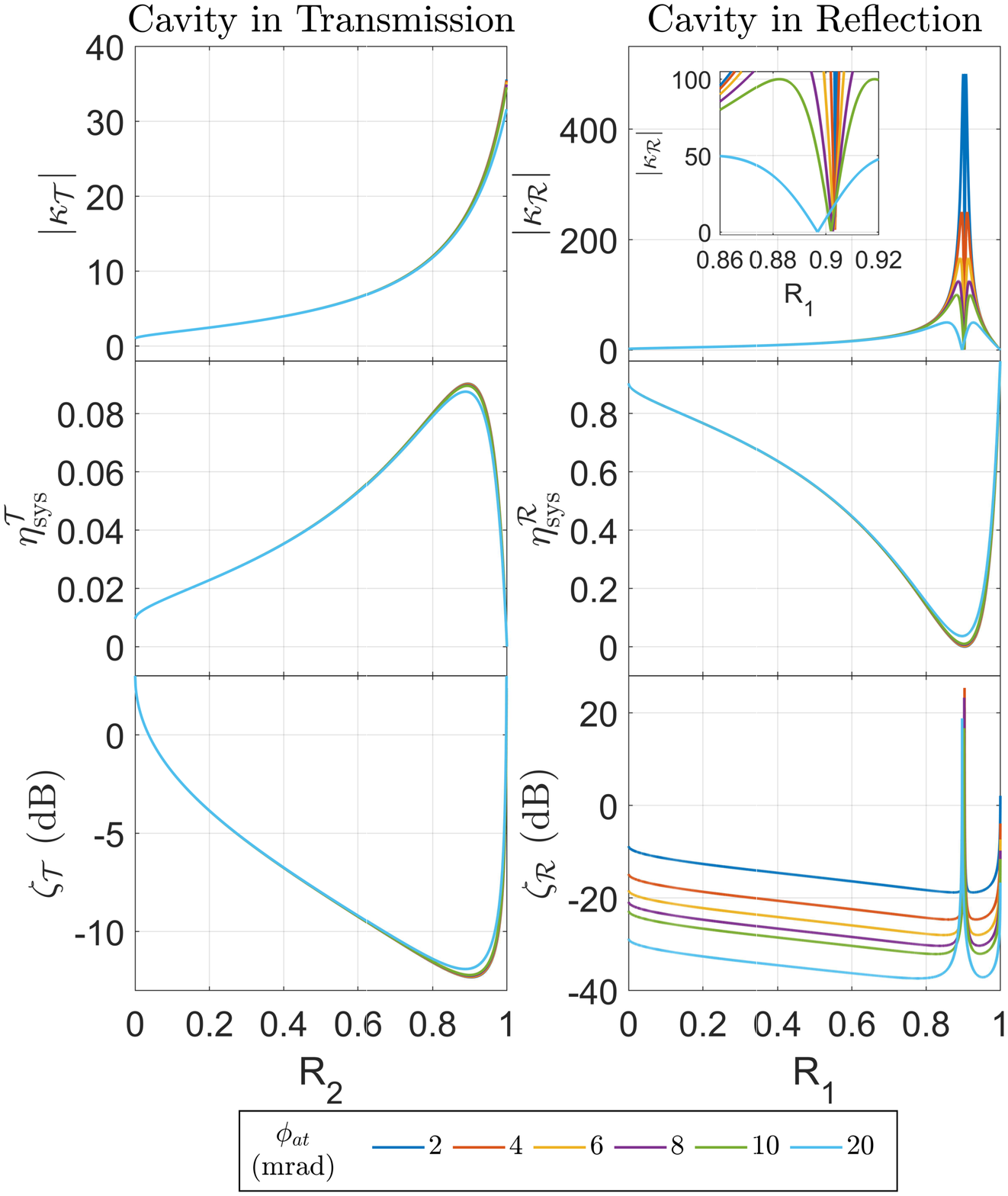}
    \caption{
    \textbf{Rotation gain, linearity and  system efficiency, for cavity-enhanced PR schemes in transmission and reflection.} Graphs show rotation gain $\kappa$, system efficiency $\eta\subsys$ and sensitivity enhancement $\zeta_{\cal R,T}$ for a range of atomic-medium rotation angles $\phiat$, as calculated with  \autoref{eq:StokesFromE}, \autoref{eq:ETransPM} and  \autoref{eq:EReflPM} in the resonant condition $\phi_0 = 0$. We take $T_G=0.975$ for the window transmission, a typical number for anti-reflection coated glass windows. 
    Left column shows transmitted signal with $R_1=0.99$, which ensures a large forward escape probability except at the highest values of $R_2$.  Right column shows reflected signal computed with $R_2 = 1$,  to maximize reflected power. } 
    \label{fig:phiat}
\end{figure}

\subsection{Results and comparison between cavity schemes}
\label{sec:Results}

\autoref{fig:phiat} shows the rotation gain $\kappa$, the system efficiency $\eta\subsys$ and sensitivity enhancement $\zeta_{\cal R,T}$  for the CE PR measurement, as a function of mirror reflectivities and for different values of $\phiat$. The $\phiat$ values studied are between $10^{-3}$ and $10^{-2}$ radians.  As one might expect, and as predicted by \autoref{eq:psicavat}, the cavity produces a strong enhancement when the $\sigma_+$ and $\sigma_-$ polarizations are simultaneously resonant with the cavity. This is possible when $\mathcal{F} \phiat  \lesssim 1$. We note that in applications to atomic media, $\phiat$ can usually be made small by choice of detuning from resonance, with no reduction in sensitivity to the rotation-producing effect, e.g., spin polarization or magnetic field. This is because increased detuning reduces both $\phiat$ and the disturbance due to inelastic scattering, in such a way that the constant-disturbance sensitivity \cite{Wolfgramm2013} is maintained  \cite{Koschorreck2010}. We consider one-sided cavities ($R_1=0.99$ for cavity in transmission or $R_2=1$ in reflection) to measure the PR from one cavity output.  The greater complexity of the reflected signal is evident.  Moreover, the reflected one shows a non-linear relationship of $\psi$ to $\phiat$, which manifests as a $\phiat$-dependent gain $\kappa_{\cal R}$, whereas $\kappa_{\cal T}$ is nearly $\phiat$-independent.

We can see that the two configurations have quite different behaviors. In reflection, $\kappa$ shows a large enhancement around a sharp, multi-lobed resonance, which moreover changes both shape and ``position,'' i.e., the $R_1$ value at which the resonance occurs, with $\phiat$. In other words, $\psi$ has a notably nonlinear relationship to $\phiat$ and this is reflected also in the sensitivity enhancement $\zeta_{\cal R}$. 
The resonance is due to destructive interference between the immediate reflection at the in-coupling mirror, which does not carry any information about the atoms, and the field exiting the cavity, which depends on $\phiat$. 

In the transmitted signal, the rotation gain grows monotonically with $R_2$ and is only weakly dependent on $\phiat$. This means $\psi$ is a nearly linear function of $\phiat$.  The system efficiency $\eta\subsys^{{\cal T}}$ shows a broad and nearly $\phiat$-independent resonance, and coincides with fairly large values of rotation gain $\kappa\subT$ and large sensitivity gain $\zeta\subT$.  

The maximum rotation gain in reflection is larger than that obtained in transmission. This apparent advantage is perhaps illusory: the strongest rotation gains occur together with the smallest efficiencies.  In this regime, the rotation angle  $\psi = \arcsin(S_2\supout/S_0\supout)$ is large because $S_0$ is small, due to the destructive interference upon reflection from a cavity near critical coupling. Such ``dark fringe'' interferometric techniques can be advantageous, especially in high-power interferometry, when the detectors' optimal power levels are below the input power \cite{LIGOCQG2015}. In probing atomic systems, however, the input power is limited by saturation of the atomic medium  \cite{Wolfgramm2013, MitchellQST2017, LuciveroPRA2017}, leading to power levels that are easily detected with typical photodetectors.  In this scenario, there is no advantage to reducing $S_0$.  

\section{Experimental implementation of cavity-enhanced polarization rotation}

\subsection{Experimental system}
\label{sec:Exp}

\begin{figure}[t]
\centering\includegraphics[scale=0.6]{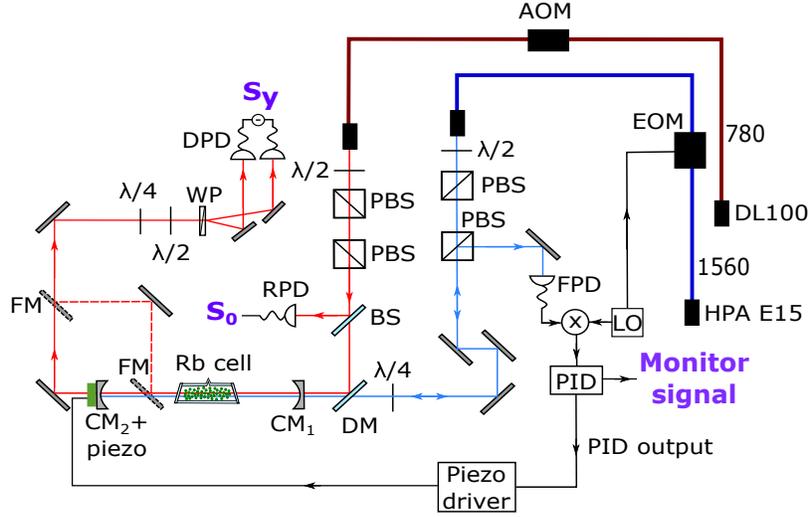}
\caption{\textbf{Experimental diagram.}  Red lines show path of \SI{780}{\nano\meter} probe light, blue lines show path of \SI{1560}{\nano\meter} locking light. Thick lines indicate propagation in fiber.  PBS - polarizing beam splitter; BS - non-polarizing beam splitter; DPD - differential photodetector; RPD - reference photodetector; FPD - fast photodiode;  DM - dichroic mirror; CM - cavity mirror; FM - flip mirror.  Dashed components indicates elements present only in the single-pass measurement. 
See text for a description of the experimental strategy. 
}\label{fig:exp_scheme}
\end{figure}

The experimental system is shown in \autoref{fig:exp_scheme}. Isotopically-enriched $^{87}$Rb vapor with no buffer gas is housed in a glass vapor cell with a single-pass length (interior dimension of the cell) of \SI{15}{\centi\meter} and with wedged windows, which are anti-reflection coated for \SI{780}{\nano\meter}. The cell is housed in a solenoid to produce an adjustable magnetic field along the cell axis (not shown in the diagram).

Cavity mirrors CM$_1$ (input) and CM$_2$ (output), each have a \SI{0.35}{\meter} radius of curvature, and are spaced to create a near-concentric cavity with an optical length of \SI{0.661(3)}{\meter}, measured from free-spectral range (FSR) measurements. CM$_2$ is mounted on a piezo-electric transducer (PZT) for precise control of the cavity length. The nominal TEM$_{00}$ mode waists at the center of the resonator are \SI{141(3)}{\micro\meter} for \SI{780}{\nano\meter} and \SI{200(4)}{\micro\meter} for \SI{1560}{\nano\meter}.  Due to the quasi-concentric geometry, higher transverse modes are separated in frequency with respect to the TEM$_{00}$ mode \cite{saleh}. In what follows, we use superscripts $p$ and $l$ to indicate probe and locking light, at wavelengths $\lambda_p = \SI{780}{\nano\meter}$ and $\lambda_l = \SI{1560}{\nano\meter}$,  respectively.  The mirrors have $R_1^{(p)}=0.990(3)$ and $R_2^{(p)}=0.860(1.5)$  and $R_1^{(l)}=0.80(2)$ and $R_2^{(l)}>0.999$ .  The measured finesses, inferred from the width and FSR observed are $\mathcal{F}^{(p)} = \SI{23\pm2}{}$ and $\mathcal{F}^{(l)} = \SI{5\pm1}{}$, implying window transmission $T_G^{(p)}= \SI{0.972\pm 0.006}{}$ and $T_G^{(l)}= \SI{0.76\pm0.06}{}$. $\mathcal{F}^{(p)}$ is measured with a photodiode placed in transmission, instead $\mathcal{F}^{(l)}$  with a photodiode placed in reflection (not shown in the schematic). A removable ``flip mirror'' inside the cavity is used to bypass CM$_2$ and thereby switch from cavity-based to single-pass measurement. 

The cavity length is stabilized using an auxiliary Pound-Drever-Hall (PDH) lock.  \SI{1560}{\nano\meter} light from a single-frequency erbium-doped fiber laser (EDFL) is phase modulated at \SI{50}{\mega\hertz} with a waveguide electro-optic modulator (EOM), circularly polarized, mode-matched to the cavity and injected into CM$_1$. The reflected power is collected on a fast photodiode and demodulated to obtain the PDH error signal. A proportional-integral-derivative (PID) controller feeds back to the PZT to maintain cavity resonance. The EDFL is frequency-doubled and its second harmonic is stabilized by modulation-transfer spectroscopy \cite{EscobarSpectroscopy} to the \textsuperscript{85}Rb D\textsubscript{2} line ($F=3 \to F^{'}=4$). 

The probe light, from a external-cavity diode laser (ECDL), is stabilized with respect to the EDFL second harmonic by an offset lock, tuneable over $\approx \SI{2}{\giga\hertz}$ around the \textsuperscript{87}Rb D\textsubscript{2} line ($F=1 \to F^{'}=2$).
An acousto-optic modulator (AOM) is used to chop the probe light into pulses. The polarization is purified with a pair of polarizing beamsplitter cubes, and half of the probe power is split off and detected on a reference photodiode (RPD) to infer $S_0\supin$. The remaining beam is spatially matched to the TEM$_{00}$ cavity mode. The locking and probe beams are shaped by independent telescopes and combined at a dichroic mirror before the cavity (not shown in the diagram). 

\begin{figure}[t]
\centering
\includegraphics[scale=0.37]{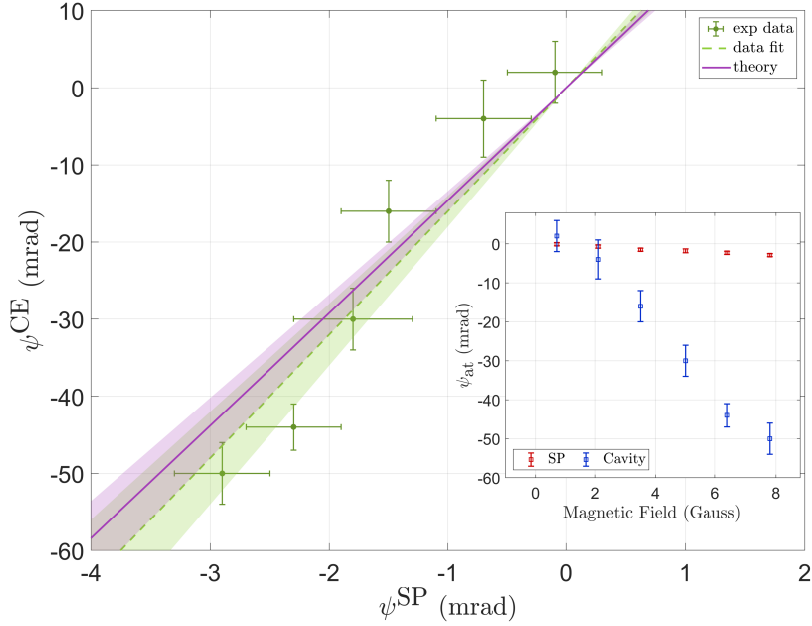}
\caption{\textbf{Measured rotation gain of cavity-enhanced probing.}  Points show the observed cavity-enhanced rotation angle $\psi\supCE$ and single-pass rotation angle $\psi\supSP$ for a variety of magnetic field strengths.  Green dashed line and green shaded region show best least-squares fit with a line passing through the origin, and plus/minus one standard error of that fit. The fit gives the experimental cavity enhancement factor $\kappa\subT^{\mathrm{exp}} = \SI{16 \pm 2}{}$. Purple line and shaded region show the model prediction $\psi\supCE = \kappa\subT \psi\supSP$ $\kappa\subT^{\mathrm{th}}$ is  calculated using  \autoref{eq:StokesFromE}, \autoref{eq:ETransPM} and  \autoref{eq:EReflPM}, with  experimental values for $R_2$ and $T_G$, and in the resonant condition $\phi_0 = 0$. This finds $\kappa\subT^{\mathrm{th}} = \SI{14.7\pm1.3}{}$, in agreement with the experimental result.  \textit{Inset}: PR angle versus magnetic field in single-pass (red squares) and in the cavity-enhanced scheme (blue squares). All error bars show plus/minus one mean standard error on 120 measurements.}\label{fig:RotationGainResults} 
\end{figure}

\subsection{Rotation gain measurements}
\label{sec:RotationGainMeas}

To measure polarization rotation, horizontally-polarized probe pulses of $\simeq \SI{20}{\micro\second}$ duration, tuned to resonance with the cavity and detuned $\simeq \SI{1.7}{\giga\hertz}$ to the red of  the $F=1 \to F'=0$ transition in $^{87}$Rb, are sent through the system. $S_2\supout$ is detected with the differential photodiode (DPD) \cite{Martin2016} and $S_0\supin$ is detected with the reference photodiode.  The PR angle is then estimated as $\hat{\psi}  =\arcsin({S_2}/{S_0})$.

Probe pulses of $\SI{1.4e7}{photons}$ were used for the single-pass probing, and $\SI{1.1e7}{photons}$ for the cavity-enhanced probing. With these numbers, the spatially-averaged probe intensity seen by the atoms is the same in the two scenarios. We note that, in the low-power, off-resonance probing regime employed here, the PR angles are negligibly affected by probe-induced nonlinearities such as saturation of the optical transition or optical pumping, so the photon number does not affect the rotation gain. Keeping the mean intensity equal is important, however, for a fair comparison of signal to noise ratios under conditions of equal illumination and thus equal perturbation to the atomic medium. 

To induce polarization rotation via the Faraday effect, a magnetic field B is applied to the atoms in a range of 0 to 10 Gauss. No optical pumping is applied to the vapor. \autoref{fig:RotationGainResults} shows a comparison of the measured rotation angles for single-pass and cavity-enhanced probing. 
The observed rotation gain is linear in the applied field and in agreement with predictions from the model of \autoref{Sec:theory}. 

With the setup coefficients given in \autoref{sec:Exp}, the parameters $\eta\subsys^{\rm SP}$ and $\eta\subsys^{\cal T}$ can be estimated. We used the definition given in \autoref{sec:SPth} and \autoref{sec:cavityTh}, where we consider $\phi_0=0$ (resonant cavity condition) and $\psi^{\rm SP} \ll 1$, in agreement to the experimental data (inset \autoref{fig:RotationGainResults}).
We obtain $\eta\subsys^{\rm SP}=\SI{0.945\pm0.012}{}$ and $\eta\subsys^{\cal T}=\SI{0.08\pm0.03}{}$. 
The parameter $\zeta\subT$ can be also calculated, with the values obtain for $\eta\subsys^{\rm SP}$ and $\eta\subsys^{\cal T}$ and with the values of $\kappa\subT^{\mathrm{exp}}$ or $\kappa\subT^{\mathrm{th}}$ written in the caption of \autoref{fig:RotationGainResults}. Using \autoref{Eq:zeta}, we have $\zeta\subT^{\rm th}=\SI{-11.8\pm1.0}{dB}$ and $\zeta\subT^{\rm exp}=\SI{-12.5\pm1.4}{dB}$, in good agreement with our prediction.

\begin{figure}[t]
\centering
\includegraphics[scale=0.5]{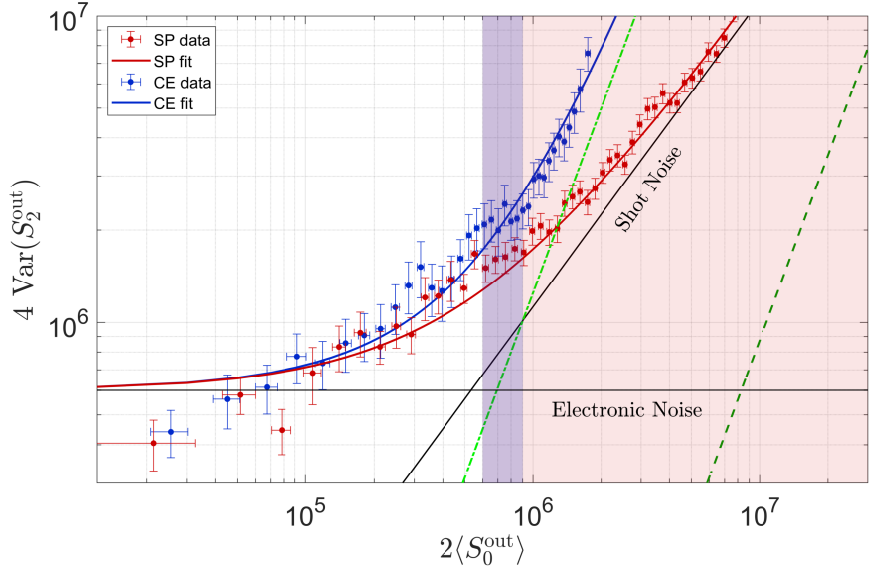}
\caption{\textbf{PR noise in the single-pass and cavity-enhanced scenarios.} Red (blue) points show  $\var(S_2\supout)$ as a function of $\braket{S_0\supout}$ in the SP (CE) scheme, as described in the text. Curves show least square fits to the data using the model in \autoref{Eq:DPDnoise}. Solid black lines show electronic and shot noise contributions common to both SP and CE scenarios. Light (dark) green dot-dashed (dashed) line shows technical noise in SP (CE) case.  Horizontal and vertical error bars show plus/minus standard error of $2 \braket{S_0\supout}$ and variance, respectively. The red (blue) shaded area marks the SN limited region of the detection system in SP (CE), which is found to be $\SI{6e5}{} \le 2 \braket{S_0\supout} \le \SI{1.3e8}{}$ ( $\SI{6e5}{} \le 2 \braket{S_0\supout} \le \SI{9e5}{}$ ).}
\label{Fig:DPDnoise}
\end{figure}

\subsection{Noise measurements} 
\label{sec:NoiseMeas}

The resonant cavity, in addition to enhancing the polarization rotation signal, can be expected to introduce additional technical noise, for example fluctuation of the cavity transmission due to fluctuations in the cavity length. To assess these noise effects, we measure the detection noise in the CE and SP cases as a function of input photon number. We follow a similar procedure to the one detailed in \cite{Martin2016}: we send light pulses of $\SI{100}{\micro\second}$ duration with different beam powers. There is no magnetic field applied on the atoms, so the expected FR angle is zero. We partition each pulse in segments of different lengths from $\SI{1}{\micro\second}$ to $\SI{100}{\micro\second}$, and extract the variance of the detected $S_2\supout$ and the mean of the detected power $S_0\supout$ (inferred from the RPD). Then we fit with the model 
\begin{equation}
\var(S_2\supout) = A + B \  \braket{S_0\supout}  + C \ \braket{S_0\supout}^2,
\label{Eq:DPDnoise}
\end{equation}
where $A,$ $B,$ and $C$ describe the strength of the electronic noise, shot noise, and technical noise contributions, respectively. \autoref{Fig:DPDnoise} shows the results  in both SP and CE scenarios. We can appreciate a clear increase of the technical noise in the CE scheme, which narrows the region of shot-noise limited detection. The CE scheme is nonetheless shot-noise limited in a range of photon numbers suitable for probing of cold atomic ensembles \cite{KubasikPRA2009}.

\section{Conclusions and outlook}
In this work we have analyzed and demonstrated cavity-enhanced polarization rotation measurements. We found general expressions, computed using the Jones calculus, for the rotation enhancement for both the forward and backward collection scenarios.  Using forward collection, we demonstrate experimentally an enhancement of $\kappa\subT^{\mathrm{exp}} = \SI{16\pm2}{}$, in good agreement with theory predictions. Noise characterization measurements show the system to be shot noise limited in a relevant range of pulse energies.  The technique is readily applicable to a variety of material systems, including hot,  cold and ultra-cold atomic and molecular ensembles, and will enable CE quantum non-demolition measurements of magnetic degrees of freedom, with potential applications in atomic magnetometers \cite{Budker2007}, gyroscopes \cite{KornackPRL2005}, and instruments to search for physics beyond the standard model \cite{LeePRL2018, gomez2020,safronova2018search}.

\begin{backmatter}
\bmsection{Funding}
This work was supported by the H2020 Future and Emerging Technologies Quantum Technologies Flagship projects MACQSIMAL (Grant Agreement No. 820393) and  QRANGE (Grant Agreement No.  820405); 
H2020 Marie Sk{\l}odowska-Curie Actions project ITN ZULF-NMR  (Grant Agreement No. 766402); PROBIST (Grant Agreement No. 754510);
Spanish Ministry of Science projects OCARINA (Grant No. PGC2018-097056-B-I00 project funded by MCIN/ AEI /10.13039/501100011033/ FEDER “A way to make Europe”), Q-CLOCKS (Grant No. PCI2018-092973 project funded by MCIN/ AEI /10.13039/501100011033/ FEDER “A way to make Europe”), and ``Severo Ochoa'' Center of Excellence CEX2019-000910-S
Generalitat de Catalunya through the CERCA program; 
Ag\`{e}ncia de Gesti\'{o} d'Ajuts Universitaris i de Recerca Grant No. 2017-SGR-1354;  Secretaria d'Universitats i Recerca del Departament d'Empresa i Coneixement de la Generalitat de Catalunya, co-funded by the European Union Regional Development Fund within the ERDF Operational Program of Catalunya (project QuantumCat, ref. 001-P-001644); Fundaci\'{o} Privada Cellex; Fundaci\'{o} Mir-Puig;
17FUN03 USOQS, which has received funding from the EMPIR programme co-financed by the Participating States and from the European Union's Horizon 2020 research and innovation programme.
This project has received funding from the European Union’s Horizon 2020 research and innovation programme under the Marie Skłodowska-Curie grant agreement No 665884.


\bmsection{Disclosures}
The authors declare no conflicts of interest.

\bmsection{Data availability} Data underlying the results presented in this paper are not publicly available at this time but may be obtained from the authors upon reasonable request.
\end{backmatter}

\bibliography{sample,MegaBib.bib,MWM.bib}

\end{document}